\documentclass[aps,twocolumn,showpacs,amsmath,amssymb]{revtex4-1}
\usepackage{graphicx}

\begin{document}

\title{Photon creation in a resonant cavity
with a nonstationary plasma mirror
\\
and its detection with Rydberg atoms}

\author{Toru Kawakubo and Katsuji Yamamoto}
\affiliation{Department of Nuclear Engineering, Kyoto University,
Kyoto 606-8501, Japan}
\date{\today}

\begin{abstract}
We investigate the dynamical Casimir effect
and its detection with Rydberg atoms.
The photons are created in a resonant cavity
with a plasma mirror of a semiconductor slab
which is irradiated by periodic laser pulses.
The canonical Hamiltonian is derived
for the creation and annihilation operators
showing the explicit time-variation in the couplings,
which originates from the external configuration
such as a nonstationary plasma mirror.
The number of created photons is evaluated as squeezing
from the Heisenberg equations with the Hamiltonian.
Then, the detection of the photons as the atomic excitations
is examined through the atom-field interaction.
Some consideration is made for a feasible experimental realization
with a semiconductor plasma mirror.
\end{abstract}


\maketitle

\section{Introduction}
\label{sec:introduction}
The quantum nature of the vacuum provides
a variety of physically interesting phenomena, including the Casimir effect
\cite{C48}.
The so-called dynamical (nonstationary) Casimir effect (DCE),
as well as the static force, has been investigated extensively
\cite{P68-69,M70,FD76,RT85,JS96,BE93-BC95-CB95,LJR96,
Law94,SPS98,D95-DK96,D98,CDM01-02,SH02,IT04-05,
Y89,LTV95,CDLM04,UPSS04,DD05-06,R06,HE06-07,RITZ09}
(and references therein), where photons are created
from the vacuum fluctuation due to nonadiabatic change of the system
such as vibration of a cavity or expansion of the universe.
It is, however, difficult experimentally to realize
the mechanical vibration of the cavity with a sufficient magnitude
at the resonant frequency $ \sim $ GHz which is required
to create a significant number of photons for detection.
As a feasible alternative, it has been proposed recently
that the oscillating wall can be simulated
by a plasma mirror of a semiconductor slab
which is irradiated by periodic laser pulses \cite{IT04-05}
(see also Refs. \cite{Y89,LTV95}).

In this paper, we investigate quantum mechanically
the photon creation via the DCE and its detection with Rydberg atoms.
We particularly intend to examine the experimental realization of DCE
with a plasma mirror of a semiconductor slab \cite{IT04-05,RITZ09}.
In Sec. \ref{sec:Hamiltonian}, the canonical Hamiltonian for DCE
is derived in terms of the creation and annihilation operators,
where the field operators are expanded simply with the initial modes.
Then, in Sec. \ref{sec:mirror} the time-varying frequencies
and squeezing couplings of the Hamiltonian are calculated
in an effective 1+1 dimensional scalar field model
with a plasma mirror.
They exhibit the enhancement of effective wall oscillation for the DCE
which is simulated by the nonstationary plasma mirror.
In Sec. \ref{sec:creation}, the number of photons created via the DCE
is evaluated as squeezing from the Heisenberg equations
for the creation and annihilation operators.
The results appear to agree essentially with those obtained
by the usual instantaneous-mode approach.
In Sec. \ref{sec:detection}, we investigate
the excitation process of Rydberg atoms through the atom-field interaction,
which is utilized to detect the created photons.
Some conditions on the physical parameters are clarified
for the efficient photon detection.
In Sec. \ref{sec:realization},
the experimental realization of DCE with a semiconductor plasma mirror
is discussed.
Section \ref{sec:summary} is devoted to a summary.

\section{Canonical Hamiltonian}
\label{sec:Hamiltonian}
We consider a scalar field in 3+1 space-time dimensions
as an effective description of the electromagnetic field
in a resonant cavity.
The Lagrangian is given by
\begin{equation}
{\cal L} = \frac{1}{2} \epsilon ({\bf x},t) ( {\dot \phi} )^2
- \frac{1}{2} ( \nabla \phi )^2
- \frac{1}{2} m^2 ({\bf x},t) \phi^2
\end{equation}
($ \hbar = c = 1 $)
\cite{Law94,SPS98,UPSS04,CDLM04,BE93-BC95-CB95}.
Here, $ \epsilon ({\bf x},t) $ and $ m^2 ({\bf x},t) $
represent the dielectric permittivity
and conductivity (effective ``mass" term), respectively,
in the matter region such as a semiconductor slab.
As specified later, they are space-time dependent,
simulating the boundary oscillation.
Conventionally, the instantaneous modes $ {\bar f}_\alpha ({\bf x},t) $
(real, orthonormal and complete) at each time $ t $
with time-varying frequencies $ {\bar \omega}_\alpha (t) $
are adopted according to the boundary oscillation:
\begin{eqnarray}
[ - \nabla^2 + m^2 ({\bf x},t) ] {\bar f}_\alpha ({\bf x},t)
= \epsilon ({\bf x},t) {\bar \omega}_\alpha^2 (t)
{\bar f}_\alpha ({\bf x},t)
\end{eqnarray}
with the orthonormalization
\begin{eqnarray}
\int_V \epsilon ({\bf x},t) 
{\bar f}_\alpha ({\bf x},t) {\bar f}_\beta ({\bf x},t) d^3 x
= \delta_{\alpha \beta} / [ 2 {\bar \omega}_\alpha (t) ] .
\end{eqnarray}
Instead, we here specify the particle representation
simply in terms of the initial modes
\begin{eqnarray}
f^0_\alpha ({\rm x}) = {\bar f}_\alpha ({\bf x},t=0) , \
\omega^0_\alpha = {\bar \omega}_\alpha (t=0) .
\end{eqnarray}
The canonical field operators in the Heisenberg picture
are expanded with the creation and annihilation operators
$ a_\alpha^\dagger (t) $ and $ a_\alpha (t) $ as
\begin{eqnarray}
\phi ({\bf x},t)
&=& \sum_\alpha [ a_\alpha (t) + a_\alpha^\dagger (t) ]
f^0_\alpha ({\bf x}) ,
\\
\Pi ({\bf x},t)
&=& \epsilon ({\bf x},0)
\sum_\alpha i \omega^0_\alpha [ - a_\alpha (t) + a_\alpha^\dagger (t) ]
f^0_\alpha ({\bf x}) ,
\end{eqnarray}
where $ \Pi ({\bf x},t) = \partial {\cal L} / \partial {\dot \phi}
= \epsilon ({\bf x},t) {\dot \phi} ({\bf x},t) $.
Then, the canonical Hamiltonian is presented by the usual procedure as
\begin{eqnarray}
H_{\rm F} (t)
&=& \int_V \frac{1}{2} \left\{ \frac{\Pi^2}{\epsilon ({\bf x},t)}
+ \phi [ - \nabla^2 + m^2 ({\bf x},t) ] \phi \right\} d^3x
\nonumber \\
&=& \sum_{\alpha} \omega_{\alpha} (t)
\left( a_\alpha^\dagger a_\alpha + \frac{1}{2} \right)
+ \sum_{\alpha \not= \beta} \mu_{\alpha \beta} (t) a_\alpha^\dagger a_\beta
\nonumber \\
&{}& + \sum_{\alpha , \beta}
i \left[ g_{\alpha \beta} (t) a_\alpha^\dagger a_\beta^\dagger
- g_{\alpha \beta}^* (t) a_\beta a_\alpha \right] ,
\label{eqn:Ht}
\end{eqnarray}
where the space-integral is taken over the whole region $ V $
which is fixed suitably (not time-dependent)
according to the physical setup,
as illustrated later for the case of a cavity
with a nonstationary plasma mirror.
[The usual oscillating boundary may also be described
as a periodic shift of the region of a high potential wall
represented by $ m^2 ({\bf x},t) $.]
The explicit time-dependence of the Hamiltonian $ H_{\rm F} (t) $
in Eq. (\ref{eqn:Ht}) represents the variation of the couplings
which originates from the nonstationary behavior
of the c-number external quantities
$ \epsilon ({\bf x},t) $ and $ m^2 ({\bf x},t) $.
The second-order field equation (Klein-Gordon equation)
is derived from the Heisenberg equations
for $ \phi ({\bf x},t) $ and $ \Pi ({\bf x},t) $.

The mode frequencies $ \omega_{\alpha} (t) $,
intermode couplings $ \mu_{\alpha \beta} (t) $
and squeezing terms $ g_{\alpha \beta} (t) $ are calculated
by considering the orthonormality of $ f^0_\alpha ({\bf x}) $
which obey the wave equation
with $ \epsilon ({\bf x},0) $ and $ m^2 ({\bf x},0) $:
\begin{eqnarray}
\omega_{\alpha} (t) &=& \omega^0_\alpha + \mu_{\alpha \alpha} (t)
\equiv \omega^0_\alpha + \delta \omega_{\alpha} (t) ,
\label{eqn:omt}
\\
\mu_{\alpha \beta} (t)
&=& 2 G^\epsilon_{\alpha \beta} (t) + 2 G^m_{\alpha \beta} (t) ,
\label{eqn:mut}
\\
g_{\alpha \beta} (t)
&=& - i [ - G^\epsilon_{\alpha \beta} (t) + G^m_{\alpha \beta} (t) ] ,
\label{eqn:gt}
\\
G^\epsilon_{\alpha \beta} (t)
&=& \frac{1}{2} \omega^0_\alpha \omega^0_\beta \int_{\delta V (t)}
\frac{\epsilon^2 ({\bf x},0)}{\epsilon_\Delta ({\bf x},t)}
f^0_\alpha ({\bf x}) f^0_\beta ({\bf x}) d^3 x ,
\label{eqn:Ge}
\\
G^m_{\alpha \beta} (t)
&=& \frac{1}{2} \int_{\delta V (t)}
m_\Delta^2 ({\bf x},t) f^0_\alpha ({\bf x}) f^0_\beta ({\bf x}) d^3 x .
\label{eqn:Gm}
\end{eqnarray}
The space-integrals for $ G^{\epsilon,m}_{\alpha \beta} (t) $
are evaluated actually in the subregion $ \delta V (t) $ 
($ \subseteq V $),
possibly time-dependent when a moving boundary is considered,
where $ \epsilon ({\bf x},t) $ and $ m^2 ({\bf x},t) $ vary in time as
\begin{eqnarray}
\epsilon_\Delta^{-1} ({\bf x},t)
& \equiv & \epsilon^{-1} ({\bf x},t) - \epsilon^{-1} ({\bf x},0) ,
\\
m_\Delta^2 ({\bf x},t) & \equiv & m^2 ({\bf x},t) - m^2 ({\bf x},0) .
\end{eqnarray}
Here, $ G^{\epsilon,m}_{\alpha \beta} (0) = 0 $
with $ \epsilon_\Delta^{-1} ({\bf x},0) = 0 $
and $ m_\Delta^2 ({\bf x},0) = 0 $ at $ t = 0 $,
as the Hamiltonian $ H_{\rm F} (0) $ is diagonalized
in terms of the initial modes $ f^0_\alpha ({\bf x}) $.

Similar formulas are presented for the effective Hamiltonian
with the instantaneous modes \cite{Law94,SPS98}.
This effective Hamiltonian involves even the time-derivatives
of the mode functions since the quantum time evolution is traced
along the instantaneous modes.
On the other hand, in the present approach the time evolution
is viewed on the initial modes according to the Heisenberg equations.
The canonical Hamiltonian is calculated
without the time-derivatives of the mode functions,
and applicable readily for various physical setups,
e.g., the case of a plasma mirror,
clarifying its dependence on the experimental parameters.
There may be some claim concerning the ambiguity
on the particle representation and photon number
since the basis modes are changing during the DCE.
This ambiguity is, however, spurious physically
(but might be essential for the case of the expanding universe,
which is beyond the present scope).
In fact, the instantaneous modes return to the initial modes
at each period of the oscillation,
where the photon number operators of the respective descriptions
coincide with each other by definition.
We can check explicitly that when the mode functions
are not deformed largely in time, as usually considered,
this canonical treatment provides essentially the same result for the DCE
as the instantaneous-mode approach.
The effects of the intermode couplings will be less significant
in the instantaneous-mode approach,
where the Hamiltonian is diagonalized at each time.
Anyway, the intermode couplings are usually off resonant,
providing subleading contributions to the DCE.

\section{Vibration with a plasma mirror}
\label{sec:mirror}
We next calculate the time-varying frequencies
and squeezing couplings of the Hamiltonian for DCE
in an effective 1+1 dimensional scalar field model
with a nonstationary plasma mirror
which is realized with a semiconductor slab irradiated
by periodic laser pulses \cite{IT04-05}.

The dielectric response of the plasma is given by
$ \epsilon ( \omega ) = \epsilon_1 [ 1 - ( \omega_p^2 / \omega^2 ) ] $
with the plasma frequency
$ \omega_p = ( n_e e^2 / \epsilon_1 m_* )^{1/2} $
in terms of the effective electron mass $ m_* $
and the conduction electron number density $ n_e $
proportional to the laser power $ W_{\rm laser}/{\rm pulse} $.
This response for the dispersion relation,
$ \epsilon ( \omega ) \omega^2
= \epsilon_1 \omega^2 - ( n_e e^2 / m_* ) $,
can be taken into account in the slab region $ [ l , l + \delta ] $
around $ x = l $ with a thickness $ \delta ( \ll L ) $ as
\begin{eqnarray}
\epsilon (x,t) = \epsilon_1 (t) ,
m^2 (x,t) = m_p^2 (t) \equiv n_e (t) e^2 / m_* ,
\label{eqn:em-xt}
\end{eqnarray}
where $ m_p^2 (0) = 0 $ for $ W_{\rm laser} (0) = 0 $.
(The spatial distribution of the conduction electrons
along the $ x $ direction may also be considered readily.)
The instantaneous mode functions are given as
\begin{equation}
{\bar f}_k (x,t) = \left\{ \begin{array}{ll}
D \sin k x & [ 0 , l ) \\
B {\rm e}^{i k^\prime x} + C {\rm e}^{- i k^\prime x}
& [ l , l + \delta ]:{\rm slab} \\
A \sin k [ x - \delta + \xi (t) ]
& ( l + \delta , L ] \end{array} \right.
\label{eqn:fkb-pm}
\end{equation}
with the dispersion relations
\begin{eqnarray}
{\bar \omega}_k^2 = ( k^2 + {\bf k}_\bot^2 ) / \epsilon_0
= ( {k^\prime}^2 + {\bf k}_\bot^2 + m_p^2 ) / \epsilon_1
\end{eqnarray}
($ k^\prime = i | k^\prime | $ for $ {k^\prime}^2 < 0 $
with large $ m_p^2 $), where $ {\bf k}_\bot $ is the momentum
in the orthogonal spatial two dimensions (not shown explicitly)
\cite{D98,CDM01-02,R06}.
The Dirichlet boundary condition
is adopted at $ x = 0 , L $ with $ \sin k [ L - \delta + \xi (t) ] = 0 $,
corresponding to the case of TE modes.
The case of TM modes can be treated in a similar way
by adopting $ m^2 (x,t)
= [ ( \partial n_e / \partial x ) e^2 / ( {\rm k}_\bot^2 m_* ) ] $
\cite{RITZ09}.

The diagonal couplings $ \delta \omega_k (t) $ and $ g_{kk} (t) $
are specifically calculated in Eqs. (\ref{eqn:omt})--(\ref{eqn:Gm})
with Eq. (\ref{eqn:fkb-pm}) for $ f^0_k (x) $ at $ t = 0 $ as
\begin{eqnarray}
\delta \omega_k (t)
&=& \omega^0_k [ \delta_\epsilon (t) + \delta_m (t) ] / L ,
\label{eqn:dom-pm}
\\
g_{kk} (t)
&=&
- (i/2) \omega^0_k [ - \delta_\epsilon (t) + \delta_m (t) ] / L .
\label{eqn:g-pm}
\end{eqnarray}
Here, the effective wall oscillation is enhanced as
\begin{eqnarray}
\delta_\epsilon (t) / \delta & \simeq &
- [ \epsilon_1 (0) / \epsilon_0 ] [ 1 - \epsilon_1 (0) / \epsilon_1 (t)]
\sin^2 kl ,
\label{eqn:dlt-e}
\\
\delta_m (t) / \delta & \simeq &
[ n_e (t) e^2 / m_* \epsilon_0 ( \omega^0_k )^2 ]
\sin^2 kl .
\label{eqn:dlt-m}
\end{eqnarray}
This effect is almost proportional to
the square of the mode function around the slab,
$ [ f^0_k (l) ]^2 \propto \sin^2 kl $,
since $ \int_l^{l+\delta} [ f^0_k (x) ]^2 dx
\simeq [ f^0_k (l) ]^2 \delta $
for $ k^\prime \delta
\sim [ \epsilon_1 (0) / \epsilon_0 ]^{1/2} ( \delta / L ) \ll 1 $
at $ t = 0 $.
If the slab is placed at the boundary $ x = l = 0 $,
$ \sin^2 kl $ is replaced
with $ ( k \delta )^2 / 3 \sim ( \delta / L )^2 \ll 1 $,
as observed in Ref. \cite{UPSS04}
claiming that the DCE is suppressed in the TE mode.
The significant photon creation, however, can take place
even in the TE mode if the slab is placed
apart from the boundaries $ x = 0 , L $
which are the nodes of $ f^0_k (x) $
\cite{CDLM04,RITZ09}.

The shift $ \xi (t) $ in the instantaneous mode
of Eq. (\ref{eqn:fkb-pm}) is determined
mainly proportional to $ \delta $
to give the frequency modulation $ \delta {\bar \omega}_k (t) $.
The diagonal squeezing coupling $ {\bar g}_{kk} (t) $ is then calculated
with the formulas for the effective Hamiltonian
\cite{Law94,SPS98}.
After some calculations we find the relations
\begin{eqnarray}
\delta {\bar \omega}_k (t) \simeq \delta \omega_k (t) , \
{\bar g}_{kk} (t) \simeq [ i / 2 {\bar \omega}_k (t) ] {\dot g}_{kk} (t) ,
\label{eqn:dom-g}
\end{eqnarray}
where the change of dielectric is assumed to be small,
$ | \epsilon_1 (t) - \epsilon_1 (0) | \ll \epsilon_1 (0) $
as usual \cite{UPSS04}.
These relations in Eq. (\ref{eqn:dom-g}) ensure
almost the same result for the DCE
in the canonical and instantaneous-mode approaches
(except for the small contribution of the off-resonant intermode couplings).
This will be checked numerically in the next section.

The above calculations of $ \delta \omega_k (t) $ and $ g_{kk} (t) $
are valid up to $ | \delta \omega_k (t) | / \omega_k^0
= | \delta_\epsilon (t) + \delta_m (t) | / L \sim 0.1 $,
which is still a significant enhancement of the effective displacement
$ | \delta_{\epsilon , m} | \gg \delta $ for the DCE.
The present approach on the fixed basis, however, does not work effectively
in an extreme situation where the mode functions are largely deformed
in time with $ | \delta \omega_k (t) | \sim \omega_k^0 $.
In such a case the instantaneous-mode approach is rather suitable
though the deformation of the mode functions
cannot be treated perturbatively \cite{RITZ09}.
Anyway, as seen in the following, a reasonable deformation
to induce $ | \delta \omega_k (t) | / \omega_k^0 \sim 0.01 - 0.1 $
is sufficient to create a significant number of photons
for detection with atoms.

\section{Photon creation as squeezing}
\label{sec:creation}
Once the Hamiltonian is presented
in terms of the creation and annihilation operators,
the time evolution for the DCE is determined by the Heisenberg equations
$ {\dot a}_\alpha (t) = i[ H_{\rm F} (t) , a_\alpha (t) ] $
and
$ {\dot a}_\alpha^\dagger (t) = i[ H_{\rm F} (t) , a_\alpha^\dagger (t) ] $.
It is described as the Bogoliubov transformation,
\begin{eqnarray}
a_\alpha (t)
&=& A_{\alpha \beta} (t) a_\beta + B_{\alpha \beta}^* (t) a_\beta^\dagger ,
\\
a_\alpha^\dagger (t)
&=& A_{\alpha \beta}^* (t) a_\beta^\dagger + B_{\alpha \beta} (t) a_\beta .
\end{eqnarray}
The master equations for the Bogoliubov transformation are derived
from the Heisenberg equations as
\begin{eqnarray}
{\dot A}_{\alpha \beta} &=& - i \omega_\alpha (t) A_{\alpha \beta}
- i \mu_{\alpha \gamma} (t) A_{\gamma \beta}
+ 2 g_{\alpha \gamma} B_{\gamma \beta} ,
\label{eqn:master-A}
\\
{\dot B}_{\alpha \beta} &=& i \omega_\alpha (t) B_{\alpha \beta}
+ i \mu_{\alpha \gamma}^* (t) B_{\gamma \beta}
+ 2 g_{\alpha \gamma}^* A_{\gamma \beta} ,
\label{eqn:master-B}
\end{eqnarray}
where the intermode couplings are renamed suitably
as $ \mu_{\alpha \gamma} ( 1 - \delta_{\alpha \gamma})
\rightarrow \mu_{\alpha \gamma} $ with $ \mu_{\alpha \alpha} \equiv 0 $.

In the following, we illustrate the characteristic features of DCE
by concentrating on a single resonant mode with time-varying frequency
$ \omega (t) = \omega_0 + \delta \omega (t) $
and squeezing coupling $ g(t) $ (the mode index ``$ k $" omitted).
The intermode couplings will not provide significant contributions
since they are fairly off resonant
generally for the nonequidistant frequency differences
\cite{D95-DK96,CDM01-02,R06}.
The master equations read
\begin{eqnarray}
{\dot A} = - i \omega (t) A + 2 g (t) B , \
{\dot B} = i \omega (t) B + 2 g^* (t) A
\label{eqn:master}
\end{eqnarray}
for the Bogoliubov transformation,
\begin{eqnarray}
a(t) = A(t) a + B^* (t) a^\dagger ,
a^\dagger (t) = A^* (t) a^\dagger + B (t) a .
\label{eqn:squeeze}
\end{eqnarray}
The solution is expressed as squeezing and phase rotation \cite{P68-69},
\begin{eqnarray}
A(t) = \cosh r(t) {\rm e}^{i \phi_A (t)} ,
B(t) = \sinh r(t) {\rm e}^{i \phi_B (t)} ,
\end{eqnarray}
with the initial condition $ A(0) = 1 , B(0) = 0 $,
ensuring $ | A(t) |^2 - | B(t) |^2 = 1 $.

An analytic solution for $ A(t) $ and $ B(t) $ is obtained
in the RWA (rotating-wave approximation) by replacing
\begin{eqnarray}
\omega (t) & \rightarrow &
\omega_0 + \langle \delta \omega \rangle ({\mbox{average}}) ,
\\
g (t) & \rightarrow & \langle g \rangle_\Omega {\rm e}^{-i \Omega t}
({\mbox{Fourier component}}) ,
\end{eqnarray}
where $ \omega_0 = \omega (0) $.
By noting the time-evolution of the number operator
$ a^\dagger (t) a(t) = | B(t) |^2 a a^\dagger + \ldots $,
we obtain the photon creation via DCE (vacuum squeezing) as
\begin{eqnarray}
n_\gamma (t) &=& \langle 0 | a^\dagger (t) a(t) | 0 \rangle = | B(t) |^2
\nonumber \\
& \simeq & \left| \frac{2 \langle g \rangle_\Omega}{\chi} \right|^2
\times \left\{ \begin{array}{ll}
\sinh^2 \chi t & ( | \Delta | < | 2  \langle g \rangle_\Omega | ) \\
| \chi |^2 t^2 & ( | \Delta | = | 2  \langle g \rangle_\Omega | ) \\
\sin^2 | \chi | t & ( | \Delta | > | 2  \langle g \rangle_\Omega | )
\end{array} \right. 
\label{eqn:ngamma}
\end{eqnarray}
with the effective squeezing rate
\begin{eqnarray}
\chi = {\sqrt{| 2 \langle g \rangle_\Omega |^2 - \Delta^2}} .
\label{eqn:chi}
\end{eqnarray}
Here, the detuning $ \Delta $ is introduced
for the frequency $ \Omega $ of laser pulses \cite{D98,CDM01-02} as
\begin{eqnarray}
\Omega = 2 ( \omega_0 + \langle \delta \omega \rangle + \Delta ) .
\label{eqn:Omega}
\end{eqnarray}
The resonance condition for DCE is then given by
\begin{eqnarray}
\Omega ({\rm resonance})
= 2 ( \omega_0 + \langle \delta \omega \rangle ) ,
\label{eqn:Omega-resonance}
\end{eqnarray}
involving the average shift of the frequency
$ \langle \delta \omega \rangle $ \cite{CDLM04,RITZ09},
rather than the naive condition $ \Omega = 2 \omega_0 $.
If $ \Omega = 2 \omega_0 $ is taken
with $ \Delta = - \langle \delta \omega \rangle $,
the squeezing rate $ \chi $ is significantly reduced,
even possibly becomes imaginary
with $ n_\gamma (t) \lesssim 1 $ oscillating as $ \sin^2 | \chi | t $.
The photon damping with the factor $ e^{- \Gamma t} $
due to the cavity loss should further be taken into account, where
\begin{eqnarray}
\Gamma = \omega_0 / Q
\end{eqnarray}
with the cavity quality factor $ Q $.
Hence, the threshold condition for the squeezing by DCE is placed as
\begin{eqnarray}
\chi > \Gamma / 2 ,
\end{eqnarray}
which is readily satisfied with a large enough $ Q $.

\begin{figure}
\includegraphics[width=.8\linewidth]{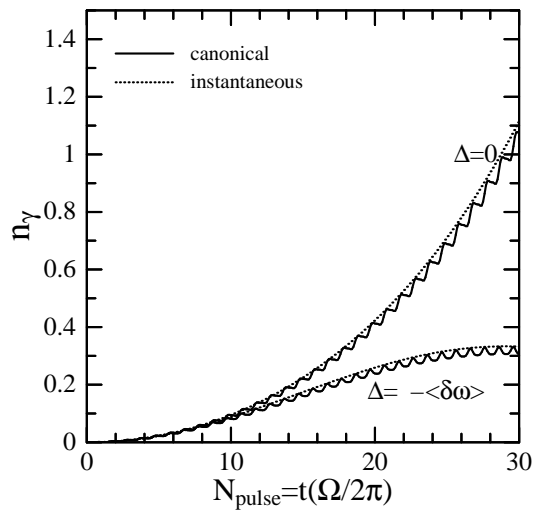}
\caption{Photon creation $ n_\gamma (t) $ (linear plot)
in the early stage of DCE
for $ N_{\rm pulse} = t ( \Omega / 2 \pi ) \leq 30 $
(the number of periodic laser pulses).
The results of the canonical and instantaneous-mode approaches
are shown with the solid and dotted curves, respectively.
The parameters are taken typically as
$ \langle \delta \omega \rangle = 0.02 \omega_0 $,
$ 2 \langle g \rangle_\Omega = i 0.01 \omega_0 $,
and $ \Delta = 0 $ (on-resonance: upper curves),
$ - \langle \delta \omega \rangle $ (off-resonance: lower curves)
for $ \Omega $.}
\label{fig:Npnps}
\end{figure}

\begin{figure}
\includegraphics[width=.8\linewidth]{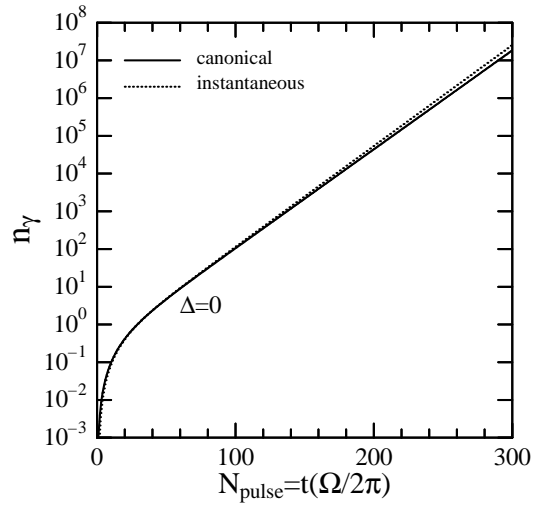}
\caption{Photon creation $ n_\gamma (t) $ (log plot)
through the DCE period
for $ N_{\rm pulse} = t ( \Omega / 2 \pi ) \leq 300 $.
The results of the canonical and instantaneous-mode approaches
are shown with the solid and dotted curves, respectively.
The parameters are taken typically as
$ \langle \delta \omega \rangle = 0.02 \omega_0 $,
$ 2 \langle g \rangle_\Omega = i 0.01 \omega_0 $,
and $ \Omega = 2.04 \omega_0 $ ($ \Delta = 0 $).}
\label{fig:Npnp}
\end{figure}

We have solved numerically the master equations
in Eq. (\ref{eqn:master}) without the RWA.
The time-varying couplings are taken typically as
$ \omega (t) = \omega_0
+ \langle \delta \omega \rangle ( 1 - \cos \Omega t ) $
and $ g(t) = 2 \langle g \rangle_\Omega ( 1 - \cos \Omega t ) $, where
$ | 2 \langle g \rangle_\Omega | \sim | \langle \delta \omega \rangle |/2 $
as indicated in Eqs. (\ref{eqn:dom-pm}) and (\ref{eqn:g-pm})
for the plasma mirror.
The instantaneous-mode solution has also been obtained
by considering the relations
$ \delta {\bar \omega} (t) = \delta \omega (t) $
and $ {\bar g} (t) = [ i / 2 {\bar \omega} (t) ] {\dot g} (t) $
in Eq. (\ref{eqn:dom-g}).
In Fig. \ref{fig:Npnps}, the photon creation $ n_\gamma (t) $
in the early stage of DCE is plotted
for $ N_{\rm pulse} = t ( \Omega / 2 \pi ) \leq 30 $
(the number of periodic laser pulses).
The results of the canonical and instantaneous-mode approaches
are shown with the solid and dotted curves, respectively.
Here, the parameters are taken typically as
$ \langle \delta \omega \rangle = 0.02 \omega_0 $,
$ 2 \langle g \rangle_\Omega = i 0.01 \omega_0 $,
and $ \Delta = 0 $ (upper curves),
$ - \langle \delta \omega \rangle $ (lower curves)
for $ \Omega $ in Eq. (\ref{eqn:Omega}).
We can see that $ n_\gamma (t) $ increases rapidly
via the DCE on the resonance
with $ \Omega = 2 ( \omega_0 + \langle \delta \omega \rangle ) $
($ \Delta = 0 $),
while $ n_\gamma (t) $ does not grow for $ \Omega = 2 \omega_0 $
($ \Delta = - \langle \delta \omega \rangle $)
due to the effective detuning brought by the average shift
$ \langle \delta \omega \rangle $.
In Fig. \ref{fig:Npnp}, the photon creation $ n_\gamma (t) $
is plotted through the DCE period
for $ N_{\rm pulse} = t ( \Omega / 2 \pi ) \leq 300 $.
The squeezing rate is determined from this plot to be
$ \chi \simeq 0.01 \omega_0 $,
as indicated in Eq. (\ref{eqn:chi}) with $ \Delta = 0 $.
This result confirms that a large number of photons
can be created via the DCE with a reasonable squeezing rate
$ \chi \sim 0.01 \omega_0 $
when the laser pulses are applied many times.
It is also found that the canonical and instantaneous-mode approaches
provide almost the same result (except for the small contribution
of the off-resonant intermode couplings).
The analytic solution under the RWA in Eq. (\ref{eqn:ngamma})
overlaps almost with the instantaneous-mode result
though it is not plotted explicitly
in Figs. \ref{fig:Npnps} and \ref{fig:Npnp}.


We briefly discuss the effect of the intermode couplings.
Specifically, the coupling
$ \mu_{12} a_1^\dagger a_2 + \mu_{12}^* a_2^\dagger a_1 $
between the modes 1 and 2 becomes resonant under a condition
$ \omega_2^0 = 3 \omega_1^0 $ $ \rightarrow $
$ \omega_2^0 - \omega_1^0 = 2 \omega_1^0 \approx \Omega $
for the case of the $ {\rm TE}_{111} $ and $ {\rm TE}_{115} $
modes in a cubic cavity due to the relation
$ ( 1^2 + 1^2 + 5^2 )^{1/2} = 3 ( 1^2 + 1^2 + 1^2 )^{1/2} $.
Then, through this resonant intermode coupling
the significant photon creation occurs
in both the modes 1 and 2 as $ n_{\gamma 1} (t) \sim n_{\gamma 2} (t) $,
increasing the total of photon numbers
\cite{CDM01-02,R06,RITZ09}.
The photons of the mode 2 are, however, fairly off resonant
with the Rydberg atoms tuned to detect the photons of the mode 1.
Hence, they cannot be detected efficiently.

\section{Detection with Rydberg atoms}
\label{sec:detection}
The photons created via the DCE are detected suitably by Rydberg atoms
with principal quantum number $ n \approx 100 $
and transition frequency $ \sim {\rm GHz} $ \cite{D95-DK96,RITZ09}.
Rydberg atoms may be treated as a two-level system
with a transition frequency $ \omega_e $
for the resonant photon absorption with $ \omega_e \approx \omega_0 $.
They are initially prepared in the lower level $ | g \rangle $,
and injected into the cavity.
A part of these atoms are excited to the upper level $ | e \rangle $
by absorbing the photons,
and detected outside the cavity as the signal of photons.
Recently, a high-sensitivity measurement of blackbody radiation
has been performed at a frequency 2.527 GHz
and low temperatures 67 mK -- 1 K
by employing a Rydberg-atom cavity detector
with a newly developed selective field ionization scheme
for $ n \approx 100 $
(the atoms excited by absorbing photons
are selectively ionized by applying an electric field) \cite{SPD06}.
Here, we note that in order to observe purely the vacuum squeezing via DCE,
the cavity should be cooled well below 100 mK
to suppress the thermal photons as $ n_\gamma ({\rm thermal}) \ll 1 $.
In fact, if photons are present initially
with an expectation value $ \langle a^\dagger a \rangle $,
they are also amplified by the DCE
as $ ( 1 + 2 | B(t) |^2 ) \langle a^\dagger a \rangle $.

Consider that $ N_{\rm Ryd} $ Rydberg atoms
(actually $ N_{\rm Ryd} \sim 100 - 1000 $ \cite{SPD06}),
which are all prepared at the lower level $ | g \rangle $, are injected
into the cavity to detect the created photons after the period of DCE,
for simplicity of argument.
(The following features for the photon detection are essentially valid
even if the atomic beam is injected continuously during and after the DCE,
as discussed later.)
The $ n_\gamma $ photons and $ N_{\rm Ryd} $ atoms
(all located at the same position for simplicity) are coupled
with the Jaynes-Cummings Hamiltonian under the RWA as
\begin{eqnarray}
H_{\rm AF} = \kappa \sqrt{N_{\rm Ryd}} ( a D_+ + a^\dagger D_- ) .
\label{eqn:HAF}
\end{eqnarray}
(The effect of the counter-rotating terms is negligible near the resonance.)
Here, the collective atomic spin-like operators are defined
(in the Schr{\"o}dinger picture) \cite {HR84} by
\begin{eqnarray}
D_+ & \equiv & \sum_{i=1}^{N_{\rm Ryd}}
| e \rangle \langle g |_{(i)} / \sqrt{N_{\rm Ryd}} ,
\label{eqn:D+}
\\
D_- & \equiv & \sum_{i=1}^{N_{\rm Ryd}}
| g \rangle \langle e |_{(i)} / \sqrt{N_{\rm Ryd}} ,
\label{eqn:D-}
\end{eqnarray}
and the complex phase for $ \kappa $ is absorbed in the atomic levels.
The single atom-photon coupling $ \kappa $ is explicitly given by
\begin{eqnarray}
\kappa = d \sqrt{\omega_0 / 2 \epsilon_0 V}
( | f^0 ({\bf x}_1 ) | / | f^0 ({\bf x}_0 ) | )
\end{eqnarray}
in terms of the magnitude of the electric dipole transition matrix element
$ d $, the cavity volume $ V $ and the mode function $ f^0 ({\bf x}) $,
where $ {\bf x}_1 $ and $ {\bf x}_0 $ represent
the atomic position and the antinode, respectively.
The collective atom-photon coupling is suitably defined by
\begin{eqnarray}
{\bar \kappa} = \kappa \sqrt{N_{\rm Ryd}} .
\end{eqnarray}
The single atom-field coupling is typically
$ \kappa \sim 3 \times 10^3 {\rm s}^{-1} $
at the antinode for the Rydberg atom of principal quantum number
$ n \approx 100 $
with $ \omega_e \approx \omega_0 \sim 1.5 \times 10^{10} {\rm s}^{-1} $
($ 2.4 {\rm GHz} \times 2 \pi $)
and $ V \sim ( 0.1 {\rm m} )^3 $ \cite{SPD06,HR84}.
Then, the collective coupling amounts
to $ {\bar \kappa} \sim 10^5 {\rm s}^{-1} \sim 10^{-5} \omega_0 $
for $ N_{\rm Ryd} \sim 10^3 $,
which is still much smaller than the resonant frequency
$ \omega_e \approx \omega_0 $.

The commutation relations among the collective operators are given by
\begin{eqnarray}
[ D_+ , D_- ] &=& D_z
\nonumber \\
& \equiv & \sum_{i=1}^{N_{\rm Ryd}}
[| e \rangle \langle e |_{(i)}-| g \rangle \langle g |_{(i)}]/N_{\rm Ryd} ,
\label{eqn:Dz}
\\
{[ D_z , D_\pm ]} &=& \pm ( 2 / N_{\rm Ryd} ) D_\pm .
\label{eqn:Dpm}
\end{eqnarray}
The operators $ {\hat N}_e $ and $ {\hat N}_g $
to represent the populations of the upper and lower levels
$ | e \rangle $ and $ | g \rangle $, respectively, are given by
\begin{eqnarray}
{\hat N}_e &=& \sum_{i=1}^{N_{\rm Ryd}} | e \rangle \langle e |_{(i)}
= ( N_{\rm Ryd} / 2 ) ( 1 + D_z ) ,
\\
{\hat N}_g &=& \sum_{i=1}^{N_{\rm Ryd}} | g \rangle \langle g |_{(i)}
= ( N_{\rm Ryd} / 2 ) ( 1 - D_z ) ,
\end{eqnarray}
satisfying the completeness
\begin{eqnarray}
{\hat N}_e + {\hat N}_g
= \sum_{i=1}^{N_{\rm Ryd}}
[ | e \rangle \langle e |_{(i)}+| g \rangle \langle g |_{(i)} ]
\equiv  N_{\rm Ryd} .
\end{eqnarray}
The created photons are detected by counting the number of excited atoms
which is represented by $ {\hat N}_e $
with eigenvalues $ 0 , 1 , \ldots , N_{\rm Ryd} $.
The initial atomic state is prepared as
\begin{eqnarray}
| 0_e \rangle = | g_1, g_2, \ldots, g_{N_{\rm Ryd}} \rangle ,
\end{eqnarray}
which is an eigenstate of $ {\hat N}_e $ with zero atomic excitation
satisfying $ D_- | 0_e \rangle = 0 $.
The one-excitation state is generated as
\begin{eqnarray}
| 1_e \rangle &=& D_+ | 0_e \rangle
\nonumber \\
&=& \frac{1}{\sqrt{N_{\rm Ryd}}} \sum_{i=1}^{N_{\rm Ryd}}
| g_1 , \ldots, e_i , g_{i+1} , \ldots ,  g_{N_{\rm Ryd}} \rangle ,
\end{eqnarray}
and so on for the multi-excitation states.

The Heisenberg equations are derived by taking
the total Hamiltonian $ H_{\rm A} + H_{\rm AF} + H_{\rm F} $
with $ H_{\rm A} = ( N_{\rm Ryd} / 2 ) \omega_e D_z $
for the free atomic system:
\begin{eqnarray}
{\dot a} &=& - i \omega_0 a - i {\bar \kappa} D_- ,
\label{eqn:a-eq}
\\
{\dot D}_- &=& - i \omega_e D_- + i {\bar \kappa} a D_z ,
\label{eqn:D-eq}
\\
{\dot D}_z &=& - i ( 2 / N_{\rm Ryd} ) {\bar \kappa}
( a D_+ - a^\dagger D_- ) .
\label{eqn:Dz-eq}
\end{eqnarray}
We solve these equations perturbatively
to see the evolution of the atomic excitation
$ N_e (t) = \langle {\hat N}_e (t) \rangle $.
First, Eqs. (\ref{eqn:a-eq}) and (\ref{eqn:D-eq})
for $ a(t) $ and $ D_- (t) = D_+^\dagger (t) $
are integrated up to the first order of $ {\bar \kappa} $
with the initial atomic operators $ D_\pm ( t_1 ) $
in Eqs. (\ref{eqn:D+}) and (\ref{eqn:D-})
and the photon operator $ a( t_1 ) $ at $ t = t_1 $
after the DCE with one sequence of $ N_{\rm pulse} $ laser pulses
for the duration
\begin{eqnarray}
t_1 = N_{\rm pulse} ( 2 \pi / \Omega ) .
\end{eqnarray}
Then, the results are applied to Eq. (\ref{eqn:Dz-eq})
to obtain $ D_z (t) $ up to the second order of $ {\bar \kappa} $.
This determines the atomic excitation as
\begin{eqnarray}
N_e (t) &=& \langle {\hat N}_e (t) \rangle
= ( N_{\rm Ryd} / 2 ) [ 1 + \langle D_z (t) \rangle ]
\nonumber \\
& \simeq & n_\gamma ( 2 {\bar \kappa} / \Delta_e )^2
\sin^2 [ \Delta_e ( t - t_1 ) / 2 ] ,
\label{eqn:net}
\end{eqnarray}
where the atomic detuning is given by
\begin{eqnarray}
\Delta_e = \omega_e - \omega_0 .
\end{eqnarray}
In these calculations, the following relations are considered:
$ \{ a , a^\dagger \} D_z + \{ D_+ , D_- \}
= 2 ( a^\dagger a D_z + D_+ D_- ) $,
$ [ a , a^\dagger ] D_z - [ D_+ , D_- ] = 0 $,
$ \langle 0_e | D_\pm ( t_1 ) | 0_e \rangle = 0 $,
$ \langle 0_e | D_+ ( t_1 ) D_- ( t_1 ) | 0_e \rangle = 0 $,
$ \langle 0_e | D_z ( t_1 ) | 0_e \rangle = - 1 $,
and $ \langle 0 | a^\dagger ( t_1 ) a ( t_1 ) | 0 \rangle = n_\gamma $
(the photons created via the DCE).
Note here that $ N_e (t) \ll N_{\rm Ryd} $
with $ \langle D_z \rangle \approx - 1 $
in the early epoch of photon detection (the linear regime).
Although it is difficult in practice to trace exactly
the time evolution beyond the linear regime
for the system of the many atoms interacting with the resonant cavity mode,
we may survey the essential features for the atomic excitation
to detect the photons as follows.

Suppose that $ n_\gamma \gg N_{\rm Ryd} $,
namely the photons are created much more than the Rydberg atoms,
as desired and feasible experimentally.
Then, the atomic excitation is eventually saturated
as $ N_e (t) \sim ({\bar \kappa} t)^2 n_\gamma \sim N_{\rm Ryd} $
for $ t \sim 1 / ( \kappa \sqrt{n_\gamma} ) $,
which is expected by extrapolating Eq. (\ref{eqn:net}) roughly
up to $ {\bar \kappa} t \sim \sqrt{N_{\rm Ryd} / n_\gamma} \ll 1 $
near the resonance $ \Delta_e \approx 0 $
(henceforth $ t - t_1 \rightarrow t $).
This excitation process may be viewed as the onset of Rabi oscillation
between $ | g \rangle $ and $ | e \rangle $ at a rate
\begin{eqnarray}
\Omega_e \sim \kappa \sqrt{n_\gamma} ,
\end{eqnarray}
which takes place almost independently for the $ N_{\rm Ryd} $ atoms
in the presence of the large field
(many photons with $ n_\gamma \gg N_{\rm Ryd} $).

On the other hand, if $ n_\gamma < N_{\rm Ryd} $
though less interesting experimentally,
the excitation is exchanged between the atoms and field
as $ N_e (t) \sim n_\gamma / 2 $ on average for $ {\bar \kappa} t \sim 1 $.
This may be understood from the fact that the interaction Hamiltonian
$ H_{\rm AF} $ in Eq. (\ref{eqn:HAF}) describes the oscillation
with a rate $ \Omega_e \sim {\bar \kappa} = \kappa \sqrt{N_{\rm Ryd}} $
between the atomic and field operators in the linear regime.
The collective atomic excitation can be treated as a quantum oscillator,
satisfying approximately the bosonic commutation relation
$ [ D_- , D_+ ] \approx - \langle D_z \rangle \approx 1 $
with $ n_\gamma \ll N_{\rm Ryd} $ in Eq. (\ref{eqn:Dz}),
that is $ D_+ $ and $ D_- $ act as the creation and annihilation operators,
respectively \cite{HR84}.

The cavity loss eventually becomes significant for $ t \gtrsim 1 / \Gamma $.
Then, the atomic excitation is also relaxed with a rate
\begin{eqnarray}
\Gamma_e \sim \left\{ \begin{array}{ll}
4 ( {\bar \kappa} / \Gamma )^2 \Gamma & ( {\bar \kappa} < \Gamma / 4 ) \\
\Gamma / 2 & ( {\bar \kappa} \geq \Gamma / 4 ) \end{array} \right.
\end{eqnarray}
through the transition $ | e \rangle \rightarrow | g \rangle + \gamma $
and the loss of the emitted photon in the cavity \cite{HR84}.
We also note that the atom-field interaction terminates
when the atoms transit through the cavity.
The atomic transit time is given by
\begin{eqnarray}
t_{\rm tr} = L/v \equiv \Gamma_{\rm tr}^{-1} ,
\end{eqnarray}
where $ v $ and $ L $ are the atomic velocity and the cavity length,
respectively.
We have typically
\begin{eqnarray}
\Gamma_{\rm tr} \sim \frac{300 {\rm m}/{\rm s}}{0.1 {\rm m}}
= 3 \times 10^3 {\rm s}^{-1} ,
\end{eqnarray}
which is comparable to the single atom-field coupling $ \kappa $.
By considering these damping effects, we realize that the created photons
are detected efficiently with the atoms under the conditions,
\begin{eqnarray}
\Omega_e & \gtrsim &  \Gamma , \Gamma_{\rm tr} ,
\\
\Gamma_{\rm tr} & \gtrsim & \Gamma_e .
\end{eqnarray}
That is, the atomic excitation should take place
for $ t \sim \Omega_e^{-1} $
before the significant loss of the created photons
due to the cavity damping ($ \Gamma \geq 2 \Gamma_e $),
and the actual cutoff of the atom-field interaction
by the atomic transit ($ \Gamma_{\rm tr} $).
It is also required that the excitation damping
($ \Gamma_e $) induced by the cavity loss
does not become significant before the atoms transit through the cavity
($ \Gamma_{\rm tr} $).

As investigated so far, if the photons are created copiously via the DCE
with $ n_\gamma \gg N_{\rm Ryd} $,
they are detected by the atomic excitation as
\begin{eqnarray}
N_e ( t_{\rm tr} ) \sim N_{\rm Ryd} / 2  .
\end{eqnarray}
Here, the condition $ \Omega_e \gtrsim \Gamma_{\rm tr} $
is less restrictive, requiring merely
$ n_\gamma \gtrsim ( \Gamma_{\rm tr} / \kappa )^2 \sim 1 $
for $ \Gamma_{\rm tr} \sim \kappa $.
The atomic detuning may be suppressed readily
as $ \Delta_e < \Omega_e $, e.g.,
for $ \Omega_e \sim 3 \times 10^6 {\rm s}^{-1} $
with $ \kappa \sim 3 \times 10^3 {\rm s}^{-1} $ and $ n_\gamma \sim 10^6 $.
The conditions $ \Omega_e \gtrsim \Gamma $
and $ \Gamma_{\rm tr} \gtrsim \Gamma_e
\sim 2 ( {\bar \kappa} / \Gamma )^2 \Gamma $
($ {\bar \kappa} < \Gamma / 4 $) imply lower and upper bounds,
respectively, on the cavity quality factor,
\begin{eqnarray}
( \omega_0 / \kappa ) / \sqrt{n_\gamma} \lesssim Q \lesssim
( \omega_0 / \kappa ) ( \Gamma_{\rm tr} / \kappa ) / N_{\rm Ryd} ,
\label{eqn:Q1}
\end{eqnarray}
where $ \omega_0 / \kappa \sim 5 \times 10^6 $.
These bounds are combined as a requirement
for the number of created photons,
\begin{eqnarray}
n_\gamma \gtrsim ( \kappa / \Gamma_{\rm tr} )^2 N_{\rm Ryd}^2
\gg N_{\rm Ryd} .
\end{eqnarray}
For example, we estimate $ Q \sim 5 \times 10^3 $ and $ n_\gamma \sim 10^6 $
for $ \Gamma_{\rm tr} \sim \kappa $ and $ N_{\rm Ryd} \sim 10^3 $.
This range of $ Q $ meets consistently the condition
$ {\bar \kappa} < \Gamma / 4 $ for $ \Gamma_e $.

On the other hand, if $ \Gamma_e = \Gamma / 2 $
($ {\bar \kappa} \geq \Gamma / 4 $)
the condition $ \Gamma_{\rm tr}\gtrsim \Gamma_e $
places a significant bound
\begin{eqnarray}
Q \gtrsim \omega_0 / \Gamma_{\rm tr} \sim 5 \times 10^6 .
\label{eqn:Q2}
\end{eqnarray}
This range of $ Q $ meets consistently
the condition $ {\bar \kappa} \geq \Gamma / 4 $ for $ \Gamma_e $.
We also note that $ N_e ( t_{\rm tr} ) \sim n_\gamma / 2 $
for $ n_\gamma < N_{\rm Ryd} $.
In this case with $ \Omega_e \sim {\bar \kappa} $,
the condition $ \Omega_e \gtrsim \Gamma $
implies $ {\bar \kappa} \geq \Gamma / 4 $.
Hence, the above range of $ Q $ in Eq. (\ref{eqn:Q2}) is effective
either for $ n_\gamma \gtrsim N_{\rm Ryd} $ or $ n_\gamma < N_{\rm Ryd} $.

The atomic beam may be injected continuously through the period of DCE.
Then, we can show that the atomic excitation is squeezed together
as $ N_e (t) \sim ( {\bar \kappa} / \omega_0 )^2 n_\gamma (t) $
during the DCE.
This atomic excitation is usually smaller
than $ N_{\rm Ryd} \sim 100 - 1000 $, e.g.,
for $ {\bar \kappa} / \omega_0 \sim 10^{-5} $ and $ n_\gamma < 10^{10} $.
Anyway, the created photons are detected with the atoms
efficiently after the DCE.

\section{Experimental realization}
\label{sec:realization}
We now discuss a feasible experimental realization
of DCE with a semiconductor plasma mirror \cite{IT04-05,RITZ09}.
Based on the analyses presented so far for the DCE and photon detection,
we can find desired values for the physical parameters.

The photons are created as
\begin{eqnarray}
n_\gamma \sim \frac{1}{4} e^{2 \chi t_1}
\sim \frac{1}{4} e^{2 \pi ( \chi / \omega_0 ) N_{\rm pulse}}
\end{eqnarray}
with the squeezing rate $ \chi $ for the resonant mode,
where $ t_1 = N_{\rm pulse} ( 2 \pi / \Omega ) $
and $ \Omega \simeq 2 \omega_0 $ (see also Fig. \ref{fig:Npnp}).
Hence, the desired number $ n_\gamma $ of created photons
places a requirement for the squeezing rate as
\begin{eqnarray}
\chi / \omega_0 \sim \frac{\ln ( 4 n_\gamma )}{2 \pi N_{\rm pulse}} .
\end{eqnarray}
Typically, $ \chi \sim 0.01 \omega_0 $
to obtain $ n_\gamma \sim 10^6 - 10^8 $
with $ N_{\rm pulse} = 300 $ laser pulses,
where the threshold condition $ \chi > \Gamma / 2 $ for the DCE
is also satisfied sufficiently with $ Q \gtrsim 10^3 $.

The effective displacement in Eq. (\ref{eqn:dlt-m})
is achieved by applying a laser power $ W_{\rm laser}/{\rm pulse} $
for the period $ T = 2 \pi / \Omega \sim 0.2 {\rm ns} $:
\begin{eqnarray}
\delta_m / L \sim ( n_{\rm s} e^2 / \epsilon_0 m_* ) L / \pi^2 ,
\end{eqnarray}
where $ \sin^2 kl = 1 $ for definiteness
(the slab is placed in the middle of cavity $ l = L/2 $),
$ \omega_0 L \sim \pi $,
and $ n_{\rm s} = n_e \delta $ ($ \propto W_{\rm laser} $)
is the surface number density of electrons.
We may readily obtain
$ ( n_{\rm s} e^2 / \epsilon_0 m_* ) L \sim 1 $ with a reasonable laser power
$ W_{\rm laser}/{\rm pulse} \sim 0.01 \mu {\rm J}/{\rm pulse} $
\cite{RITZ09},
achieving a significant displacement $ \delta_m \sim 0.1 L $.
In this case, the conductivity effect $ \delta_m $ in Eq. (\ref{eqn:dlt-m})
dominates over the dielectric effect $ \delta_\epsilon $
in Eq. (\ref{eqn:dlt-e}) for $ \epsilon_1 (0) \sim 1 - 10 $
and $ \epsilon_1 (0) \leq | \epsilon_1 (t) | $
[the photon damping by the complex $ \epsilon_1 (t) $
does not exceed the squeezing by the DCE mainly with $ \delta_m $].
We estimate the variation of the mode frequency as
\begin{eqnarray}
\delta \omega \simeq ( \delta_m / L ) \omega_0
\sim 0.1 \omega_0 ( W_{\rm laser} / 0.01 \mu {\rm J} ) .
\end{eqnarray}
By noting the relation $ | \delta \omega | \simeq | 2 g | $,
the desired squeezing rate for the DCE can be obtained
in Eq. (\ref{eqn:chi}) with $ \Delta = 0 $ as
\begin{eqnarray}
\chi = | 2 \langle g \rangle_\Omega |
\sim 0.01 \omega_0
( r_\Omega / 0.1 ) ( W_{\rm laser} / 0.01 \mu {\rm J} ) .
\end{eqnarray}
Here, the factor $ r_\Omega $ represents the Fourier component
$ \langle g \rangle_\Omega {\rm e}^{- i \Omega t} $ of $ g(t) $,
which may be optimized by suitably designing the time-profile
$ W_{\rm laser} (t) $ of laser pulse.
As seen in Eqs. (\ref{eqn:Omega}) and (\ref{eqn:Omega-resonance}),
the tuning of $ \Omega $ is required for the resonance
by taking into account the average shift
$ \langle \delta \omega \rangle / \omega_0 \sim 0.01 - 0.1 $.

As for the photon detection, the analyses in Sec. \ref{sec:detection}
indicate that roughly $ N_{\rm Ryd}/2 \sim 100 $ atomic excitations
are detected per mean atomic transit time $ t_{\rm tr} \sim 0.1 {\rm ms} $
for the creation of $ n_\gamma \sim 10^6 - 10^8 $ photons via the DCE.
The quality factor of cavity should be chosen suitably
to ensure the efficient atomic excitation and detection.
Specifically, $ Q \sim 5 \times 10^3 $ in Eq. (\ref{eqn:Q1})
or $ Q \gtrsim 5 \times 10^6 $ in Eq. (\ref{eqn:Q2}).
We note that even if an excessive amount of photons
($ n_\gamma \gg 10^8 $) are created,
their detection is actually limited by the number of Rydberg atoms
$ N_{\rm Ryd} \sim 100 - 1000 $.
After the detection, the photons remaining in the cavity are relaxed finally
as $ n_\gamma \rightarrow 0 $
for $ t \gtrsim 10 {\rm ms} \gg \Gamma^{-1} , t_{\rm tr} $;
namely the field returns to the vacuum.
Then, the subsequent rounds of photon creation and detection
are performed repeatedly.

\section{Summary}
\label{sec:summary}
We have investigated quantum mechanically
the photon creation via DCE and its detection with Rydberg atoms,
specifically considering the experimental realization
in a resonant cavity with a plasma mirror of a semiconductor slab
irradiated by laser pulses.
The canonical Hamiltonian for the DCE is derived
in terms of the creation and annihilation operators
showing the explicit time-variation
which originates from the external configuration
such as the nonstationary plasma mirror.
Then, the photon creation is evaluated as squeezing
from the Heisenberg equations.
This confirms that a sufficiently large number of photons can be created
via the DCE with a reasonable squeezing rate
when the laser pulses are applied many times.
The atomic excitation process to detect the photons
is described with the atom-field interaction,
which clarifies the conditions for the efficient detection.
Based on these analyses, desired values of the physical parameters
are considered for a feasible experiment for DCE and its detection
with a plasma mirror and Rydberg atoms.

\begin{acknowledgments}
The authors appreciate valuable discussions
with S. Matsuki, Y. Kido, T. Nishimura, W. Naylor
and the Ritsumeikan University group.
\end{acknowledgments}

\end{document}